\documentstyle[multicol,pra,aps,psfig]{revtex}

\begin{document}
\title{Continuous-variable quantum teleportation through lossy channels}
\author{A.V. Chizhov%
\footnote{Permanent address:
Joint Institute for Nuclear Research,
Bogoliubov Laboratory of Theoretical Physics,
141980 Dubna, Moscow Region, Russia},
L. Kn\"oll, and D.-G. Welsch}
\address{Friedrich-Schiller-Universit\"{a}t Jena,
Theoretisch-Physikalisches Institut
\\
Max-Wien-Platz 1, D-07743 Jena, Germany}

%\date{\today}
\maketitle

\begin{abstract}
The ultimate limits of continuous-variable single-mode quantum
teleportation due to absorption are studied, with special emphasis
on \mbox{(quasi-)}mo\-nochromatic optical fields propagating through fibers.
It is shown that even if an infinitely squeezed two-mode squeezed
vacuum were used, the amount of information that would be
transferred quantum mechanically over a finite distance
is limited and effectively approaches to zero
on a length scale that is much shorter than the (classical)
absorption length. Only for short distances the
state-dependent teleportation fidelity can be close
to unity. To realize the largest possibly fidelity,
an asymmetrical equipment must be used, where the
source of the two-mode squeezed vacuum is nearer
to Alice than to Bob and in consequence the coherent displacement
performed by Bob cannot be chosen independently
of the transmission lengths.
\end{abstract}
\vspace*{0.4cm}
\hspace*{1.4cm}
PACS number(s): 03.67.-a, 42.50.Dv, 42.79.-e

\begin{multicols}{2}

%%%%%%%%%%%%%%%%%%%%%%%%%%%%%%%%%%%%%%%%%%%%%%%%%%%%%%%%%%%%%%%%%%%%%%%%%%%%%

\section{Introduction}
\label{intro}

Quantum teleportation, in which an unknown quantum
state is teleported from a sending station to a
distant receiving station, has been one of the exciting
manifestations of quantum-state entanglement of bipartite
systems. Schemes for both spin-like quantum states
\cite{Benn93,Sten98} and continuous-variable quantum states
\cite{Vaid94,Brau98,Milb99,Hor00,Loock00,Clausen00} have been
proposed, and experiments have been performed
\cite{Bouw97,Boschi98,Ralph98,Furu98}.
The very idea of quantum teleportation is to transfer
that part of information on the (unknown) state which is
lost in a single measurement quantum mechanically by means
of appropriately entangled states.

In continuous-variable teleportation the sender (Alice) and
the recipient (Bob) must share a highly entangled
state in order to be able to really teleport an
{\em arbitrary} quantum state. For teleporting a single-mode
quantum state, a two-mode squeezed vacuum (TMSV)
is commonly assumed to play the role of the entangled state.
High entanglement then means high squeezing, which implies
an entangled macroscopic (at least mesoscopic) state.
However, entanglement is known to sensitively respond to
environment influences, which unavoidably give rise to
entanglement degradation \cite{Duan97,Scheel00a,Scheel00} and
thus reduce the fidelity of teleportation, as was shown
in Ref.~\cite{Kim00}, where the two modes were equally coupled
to some heat bath.

The aim of the present paper is to study the ultimate
limits of quantum teleportation that arise from absorption
during the propagation of the two modes from the source
of the TMSV to Alice and Bob, so that they have one each
for further manipulation. With regard to optical fields
that are desired to propagate over longer distances, fibers
would preferably be used. As we will see,
the ratios of the propagation length to the low-temperature
absorption length essentially determine the
amount of quantum-mechanically transferable information.
In this way, the fidelity of teleportation becomes not only
state-dependent, but also dependent on the position of the
TMSV source relative to the positions of Alice and Bob.
Thus, the original concept of teleportation of a really unknown
quantum state to a really distant position becomes questionable.

The paper is organized as follows. Section \ref{sec2} presents the
basic equations, with special emphasis on the entangled state that
is shared by Alice and Bob in practice and Bob's choice of
the displacement after Alice's measurement. In Section
\ref{sec3} the theory is applied to the teleportation of
squeezed states and number states and a detailed analysis of
the various dependencies are given. Finally, some concluding
remarks are given in Section \ref{sec4}.

%%%%%%%%%%%%%%%%%%%%%%%%%%%%%%%%%%%%%%%%%%%%%%%%%%%%%%%%%%%%%%%%%%%%%%%%%%%%%%
%%%%%%%%%%%%%%%%%%%%%%%%%%%%%%%%%%%%%%%%%%%%%%%%%%%%%%%%%%%%%%%%%%%%%%%%%%%%%%

\section{Basic equations}
\label{sec2}

In what follows we consider the standard scheme of continuous-variable
single-mode teleporation, assuming the entangled state is a (strongly)
squeezed TMSV. One mode is transmitted to Alice (sender)
and the other one to Bob (recipient). Since the transmission, e.g.,
through fibers is unavoidably connected with some losses, the state
effectively shared by Alice and Bob is not the originally generated
TMSV but a mixed state, whose entanglement
drastically decreases with the distance between Alice and Bob
\cite{Scheel01}.

%%%%%%%%%%%%%%%%%%%%%%%%%%%%%%%%%%%%%%%%%%%%%%%%%%%%%%%%%%%%%%%%%%%%%%%%%%%%%%

\subsection{The teleported state}
\label{sec2.1}

Let us briefly repeat the main stages of teleportation.
If $W_{\rm in}(\gamma)$ is the Wigner function of the
signal-mode quantum state that is desired to be teleported
and $W_{\rm out}^{\rm E}(\alpha,\beta)$ is the Wigner function
of the entangled state that is effectively shared by Alice and Bob,
the Wigner function of the (three-mode) overall system then reads
\begin{equation}
\label{2.1}
W(\gamma,\alpha ,\beta ) = W_{\rm in}(\gamma)
\,W_{\rm out}^{\rm E}(\alpha,\beta).
\end{equation}
After combination of the signal mode and Alice's mode of the
entangled two-mode system through a 50\%:50\% (lossless) beam
splitter the Wigner function changes to
\begin{equation}
\label{2.2}
W(\mu ,\nu ,\beta )
   = W_{\rm in}\!\left( \frac{\mu -\nu }{\sqrt{2}} \right)
   W_{\rm out}^{\rm E}\!
   \left( \frac{\mu +\nu }{\sqrt{2}}, \beta \right).
\end{equation}

Measurement of the real part of $\mu $, $\mu _{\rm R}$,
and the imaginary part of $\nu $, $\nu _{\rm I}$,
then prepares Bob's mode in a quantum state whose
Wigner function is given by
\begin{eqnarray}
\label{2.3}
\lefteqn{
W(\beta |\mu _{\rm R},\nu _{\rm I}) = \frac{1}{P(\mu _{\rm R},\nu _{\rm I})}
}
\nonumber\\[.5ex]&&\hspace{0ex}\times
   \int d\nu _{\rm R} \int d\mu _{\rm I} \,
   W_{\rm in}\!\left( \frac{\mu -\nu }{\sqrt{2}} \right)
   W_{\rm out}^{\rm E}\!\left( \frac{\mu +\nu }{\sqrt{2}}, \beta \right) ,
\end{eqnarray}
where
\begin{equation}
\label{2.4}
P(\mu _{\rm R},\nu _{\rm I})
   = \int d\nu _{\rm R} \int d\mu _{\rm I} \int d^2\beta
   \, W(\mu ,\nu ,\beta )
\end{equation}
is the probability
density of measuring $\mu _{\rm R}$ and $\nu _{\rm I}$.
Introducing the complex variables
\begin{equation}
\label{2.5}
\gamma = \left(\mu - \nu \right)/\sqrt{2},
\quad
\gamma ' =\sqrt{2} \left( \mu _{\rm R} - i \nu _{\rm I} \right),
\end{equation}
we may rewrite Eq.~(\ref{2.3}) as
\begin{equation}
\label{2.6}
W(\beta |\gamma ')
   = \frac{1}{P(\gamma ')}
   \int d^2\gamma \,W_{\rm in}(\gamma ) \, W_{\rm out}^{\rm E}
   ({\gamma '}^{*} \!-\! {\gamma }^{*}, \beta )
\end{equation}
[$P(\mu_{\rm R},\nu_{\rm I})/2$ $\!\to$ $\!P(\gamma')$].

Depending upon the result of Alice's measurement,
Bob now coherently displaces the quantum state of his mode
in order to generate a quantum state whose  Wigner function is
\mbox{$W(\beta$ $\!-$ $\!\Delta(\gamma')|\gamma ')$}.
If we are not interested
in the one or the other measurement result, we may average
over all measurement results to obtain the teleported
quantum state on average:
\begin{eqnarray}
\label{2.7}
\lefteqn{
W_{\rm out}(\beta) =
   \int d^2\gamma'\, P(\gamma')
\,
W\!\left(\beta  - \Delta (\gamma ')|\gamma'\right)
}
\nonumber\\[.5ex]\hspace{0ex}&&
= \int d^2\gamma \,W_{\rm in}(\gamma ) \!\int d^2 \gamma'
   \,W_{\rm out}^{\rm E}\!\left({\gamma '}^{*} \!-\! {\gamma }^{*},
   \beta \!-\! \Delta (\gamma ')\right) .
\end{eqnarray}

%%%%%%%%%%%%%%%%%%%%%%%%%%%%%%%%%%%%%%%%%%%%%%%%%%%%%%%%%%%%%%%%%%%%%

\subsection{Available entangled state}
\label{sec2.2}

Let us assume that the modes of the originally generated
TMSV propagate to Alice and Bob through
fibers of (spectral) transmission coefficients $T_1(\omega)$ and
$T_2(\omega)$, respectively. When
\begin{eqnarray}
\label{2.8}
\lefteqn{
W_{\rm in}^{\rm E}(\alpha,\beta) =
   \frac{4}{\pi ^2}\, \exp\!\left[ -2 \left( |\alpha|^2
   +|\beta|^2 \right) \cosh|2\zeta|
\right.
}
\nonumber\\[.5ex]&&\hspace{6ex}
   + \left. 2 \left( e^{-i\varphi} \alpha\beta
   + e^{i\varphi } \alpha ^* \beta ^*  \right) \sinh|2\zeta | \right]
\end{eqnarray}
is the Wigner function of the originally generated TMSV
(\mbox{$\zeta$ $\!=$ $\!|\zeta |e^{i\varphi}$}, squeezing parameter), then
the Wigner function of the quantum state the two modes are prepared in after
transmission takes the form of \cite{Duan97,Scheel00,Chizh00}
\begin{eqnarray}
\label{2.9}
\lefteqn{
W_{\rm out}^{\rm E}(\alpha,\beta) =  \frac{4}{\pi ^2 \cal{N}}
}
\nonumber\\[.5ex]&&\hspace{2ex}\times\,
   \exp\!\left[- 2 \bigl(C_2|\alpha|^2+C_1|\beta|^2
   + S^*\alpha\beta+S\alpha ^* \beta ^* \bigr) \right],
\end{eqnarray}
where ($i$ $\!=$ $\!1,2$)
\begin{equation}
\label{2.10}
S=\frac{e^{i\varphi }}{\cal{N}}\,
   T_1 T_2
   \sinh|2\zeta|,
\end{equation}
\begin{equation}
\label{2.11}
C_i = \frac{1}{\cal{N}}
   \left[ 1 + |T_i|^2 \left(\cosh|2\zeta |-1\right)
   +\, 2n_{{\rm th} \, i}\left( 1-|T_i|^2\right)\right],
\end{equation}
\begin{eqnarray}
\label{2.12}
\lefteqn{
{\cal N} =
   \left[ 1 + |T_1|^2 \left(\cosh|2\zeta |-1\right)
   + 2n_{{\rm th} \, 1}\left( 1-|T_1|^2\right) \right]
}
\nonumber\\[.5ex]&&\hspace{3ex}\times\,
   \left[ 1 + |T_2|^2 \left(\cosh|2\zeta |-1\right)
   +2n_{{\rm th} \, 2}\left( 1-|T_2|^2\right) \right]
\nonumber\\[.5ex]&&\hspace{3ex}
   - \,|T_1 T_2|^2 \sinh ^2 |2\zeta| ,
\end{eqnarray}
with $n_{{\rm th} \, i}$ $\!=$ $\!\{\exp[\hbar\omega/
(k_{\rm B}\vartheta_i)]$ $\!-$ $\!1\}^{-1}$ being the mean number of
thermal excitations at temperature $\vartheta_i$.
It may be useful to
express $|T_i|$ in terms of the ratio of the transmission
length $l_i$ to the absorption length $l_{{\rm A}\, i}$ such that
\begin{equation}
\label{2.14}
|T_i|=\exp (-l_i/l_{{\rm A} \, i} ).
\end{equation}

It should be pointed out that Eq.~(\ref{2.9})
together with Eqs.~(\ref{2.10}) -- (\ref{2.12})
directly follows from the general formalism of quantum-state
transformation at absorbing four-port devices
\cite{Grun96,Knoell99} for
vanishing reflection coefficients.
For nonvanishing reflection
coefficients, the terms
\mbox{$n_{{\rm th} \, i}(1$ $\!-$ $\!|T_i|^2)$}
in Eqs.~(\ref{2.11}) and (\ref{2.12}) must be simply replaced with
\mbox{$n_{{\rm th} \, i}(1$ $\!-$ $\!|T_i|^2$ $\!-$ $\!|R_i|^2)$}.

%%%%%%%%%%%%%%%%%%%%%%%%%%%%%%%%%%%%%%%%%%%%%%%%%%%%%%%%%%%%%%%%%%%%%%%

\subsection{Fidelity}
\label{sec2.3}

Let us assume that the quantum state to be teleported is a
pure one, \mbox{$\hat{\varrho}_{\rm in}$ $\!=$
$\!|\psi _{\rm in}\rangle \langle \psi _{\rm in}|$}.
A measure of how close to it is the (mixed) output quantum state
$\hat{\varrho}_{\rm out}$ may be the teleportation fidelity
\begin{equation}
\label{2.15}
F = \langle \psi_{\rm in}| \hat{\varrho}_{\rm out} |\psi _{\rm in}\rangle \,.
\end{equation}
Using the well-known representation of the density operator in terms
of the coherent displacement operator $\hat{D}(\xi)$
\cite{Cahill69,Glauber69}
\begin{equation}
\label{2.16}
\hat{\varrho}
   = \frac{1}{\pi} \int d^2 \xi\, \chi (\xi ) \hat{D}^{\dagger} (\xi ),
\end{equation}
with $\chi (\xi )$ being the Fourier transform of the Wigner function,
Eq.~(\ref{2.15}) can be rewritten as
\begin{equation}
\label{2.17}
F = \frac{1}{\pi} \int d^2 \xi\, \chi _{\rm in} (\xi )
   \chi^*_{\rm out} (\xi ).
\end{equation}
Equivalently, the fidelity can be given by the overlap
of the Wigner functions:
\begin{equation}
\label{2.18}
F = \pi \int d^2 \beta\, W_{\rm in} (\beta ) W_{\rm out} (\beta ).
\end{equation}

Perfect teleportation implies unity fidelity;
that is perfect overlap of the Wigner functions of the
input and the output quantum state. Clearly, losses
prevent one from realizing this case, so that the really
observed fidelity is always less than unity. Thus,
the task is to choose the scheme-inherent parameters
such that the fidelity is maximized.

%%%%%%%%%%%%%%%%%%%%%%%%%%%%%%%%%%%%%%%%%%%%%%%%%%%%%%%%%%%%%%%%%%%%%%%

\subsection{Choice of the displacement}
\label{sec2.4}

An important parameter that must be specified is the displacement
\mbox{$\beta$ $\!\to$ $\!\beta$ $\!-$ $\!\Delta (\gamma ' )$}
[in Eq.~(\ref{2.7})], which has to be performed by Bob after Alice's
measurement. For this purpose, we substitute Eq.~(\ref{2.9})
into Eq.~(\ref{2.6}) to obtain, on using the relation
\mbox{$C_1C_2$ $\!-$ $\!|S|^2$ $\!=$ $\!{\cal N}^{-1}$},
\begin{eqnarray}
\label{2.19}
\lefteqn{
W(\beta |\gamma ') = \frac{1}{P(\gamma ')}
   \frac{2}{\pi {C_2\cal{N}}}\,
   \exp\!\left( - \frac{2}{C_2\cal{N}} |\beta |^2 \right)
}
\nonumber\\[.5ex]&&\hspace{2ex}\times
   \int d^2\gamma\, \frac{2C_2}{\pi }\,
   \exp\!\left( \!-2C_2\left| \gamma ' \!-\! \gamma
   \!+\!\frac{S^*}{C_2} \beta \right| ^2 \right)
   W_{\rm in} (\gamma ).
\end{eqnarray}
Here we have restricted our attention to optical fields
whose thermal excitation may be disregarded
(\mbox{$n_{{\rm th}\,i}$ $\!\approx$ $\!0$}).
{F}rom Eqs.~(\ref{2.11}) and (\ref{2.12}) it follows that,
for not too small values of the (initial) squeezing parameter
$|\zeta|$, the
variance of the Gaussian in the first line
of Eq.~(\ref{2.19}),
$C_2{\cal{N}}/4$,
increases with $|\zeta|$ as
$e^{2|\zeta|}|T_2|^2/8$,
whereas the
variance of the Gaussian in the integral in the second line,
$1/(4C_2)$,
rapidly approaches the (finite) limit
($T_2$ $\!\not=$ $\!0$)
\begin{equation}
\label{2.20}
\sigma _{\infty}= \lim_{|\zeta|\to\infty}
   \frac{1}{4C_2}
   = \frac{|T_1|^2 + |T_2|^2 - 2 |T_1 T_2|^2}
    {4 |T_2|^2}\,.
\end{equation}

Thus, Bob's mode is prepared (after Alice's
measurement) in a quantum state that is obtained,
roughly speaking, from the input quantum state by
shifting the Wigner function according to \mbox{$\gamma$
$\!\to$ $\!\gamma'$ $\!+$ $\!\beta S^*/C_2$} and smearing
it over an area whose linear extension is given by
$2\sqrt{\sigma_\infty}$.
It is therefore expected that the best what Bob can do is to
perform a displacement with
\begin{equation}
\label{2.21}
\Delta (\gamma') = e^{i\tilde{\varphi} } \lambda \gamma',
\end{equation}
where $\tilde{\varphi}$ $\!=$
$\!\varphi$ $\!+$ $\!\arg T_1$ $\!+$ $\arg T_2$, and
\begin{equation}
\label{2.23}
\lambda  = \lim_{|\zeta|\to\infty} \frac{C_2}{|S|}
= \left| \frac{T_2}{T_1} \right| .
\end{equation}
Substitution of this expression into Eq.~(\ref{2.7}) yields
\begin{equation}
\label{2.22}
\hspace{-.5ex}
W_{\rm out}(\beta e^{i\tilde{\varphi}})
= \frac{1}{2\pi\sigma\lambda^2}
\!\!\int \!d^2 \gamma \, W_{\rm in}(\gamma )
\exp\!\left(\! -\frac{|\gamma \!-\! \beta/\lambda|^2}{2\sigma }
\!\right)\!,
\end{equation}
where
\begin{equation}
\label{2.24}
\sigma = \frac{{\cal{N}}}{4 \lambda ^2}
\left( C_2 + \lambda^2 C_1 - 2 \lambda |S| \right).
\end{equation}
Note that $\lim_{|\zeta|\to\infty}\sigma$ $\!=$ $\!\sigma_\infty$.

Clearly, even for arbitrarily large
squeezing, i.e, \mbox{$|\zeta|$ $\!\to$ $\infty$}, and thus
arbitrarily large entanglement, the input quantum state
cannot be scanned precisely due to the unavoidable losses,
which drastically reduce the amount of information that can be
transferred nonclassically from Alice to Bob.
Let $\delta_W$ be a measure of the (smallest) length scale
in phase space on which
the Wigner function of the signal-mode state,
$W_{\rm in}(\gamma)$, typically changes. Teleportation
then requires, apart from the scaling by $\lambda$, that
the condition
\begin{equation}
\label{2.25}
\sigma _{\infty} \ll \delta _W^2
\end{equation}
is satisfied. Otherwise, essential information about
the finer points of the quantum state are lost.
For given $\delta _W$, the condition (\ref{2.25}) can be
used in order to determine the ultimate limits of teleportation,
such as the maximally possible distance between Alice and
Bob. In this context, the question of the optimal position
of the source of the TMSV arises. Needless to say, that all the
results are highly state-dependent.

%%%%%%%%%%%%%%%%%%%%%%%%%%%%%%%%%%%%%%%%%%%%%%%%%%%%%%%%%%%%%%
%%%%%%%%%%%%%%%%%%%%%%%%%%%%%%%%%%%%%%%%%%%%%%%%%%%%%%%%%%%%%%

\section{Squeezed and number states}
\label{sec3}

Let us illustrate the problem for
squeezed and number states. Applying the general formulas
given in Section \ref{sec2} to these classes of states,
all calculations can be performed analytically and closed
expressions for the fidelity can be derived. They will enable us,
to see the effect of the displacement and the position
of the TMSV source in more detail.

%%%%%%%%%%%%%%%%%%%%%%%%%%%%%%%%%%%%%%%%%%%%%%%%%%%%%%%%%%%%%%%

\subsection{Squeezed states}
\label{sec3.1}

Let us first assume that the unknown single-mode quantum state,
which is desired to be teleported, is a squeezed coherent state.
Its Wigner function can be given by
\begin{eqnarray}
\label{3.1}
\lefteqn{
W_{\rm in}(\gamma ) =\frac{N_{\rm in}}{\pi }
\,\exp\!\left[ - A_{\rm in}|\gamma |^2
\right.
}
\nonumber\\&&\hspace{2ex}
\left.
-\,B_{\rm in}\left(\gamma ^2 +  \gamma ^{\ast 2}\right)
+ C_{\rm in}^\ast\gamma  + C_{\rm in}\gamma^\ast\right],
\end{eqnarray}
where
\begin{eqnarray}
\label{3.2}
\lefteqn{
N_{\rm in} = 2 \exp\!\left[-2 |\alpha _0|^2\cosh\!\left(2\zeta _0\right)
\right.
}
\nonumber\\&&\hspace{6ex}
\left.
- \left(\alpha _0^2 +\alpha _0^{*2} \right)
\sinh\!\left( 2\zeta _0\right) \right],
\end{eqnarray}
\begin{equation}
\label{3.3}
A_{\rm in} = 2\cosh(2\zeta _0),
\end{equation}
\begin{equation}
\label{3.4}
B_{\rm in} = \sinh(2\zeta _0),
\end{equation}
\begin{equation}
\label{3.5}
C_{\rm in} = 2 \left[ \alpha _0 \cosh \!\left(2\zeta _0\right)
+ \alpha _0^* \sinh \!\left(2\zeta _0\right) \right].
\end{equation}
Here, $\alpha_0$ is the coherent amplitude and $\zeta_0$ is the
squeezing parameter, which is chosen to be real.
Substituting Eq.~(\ref{3.1}) [together with Eqs.~(\ref{3.2}) --
(\ref{3.5})] into Eq.~(\ref{2.22}), we derive
(Appendix \ref{app1})
\begin{eqnarray}
\label{3.6}
\lefteqn{
W_{\rm out}(\beta ) =\frac{N_{\rm out}}{\pi }
\,\exp\!\left[ - A_{\rm out}|\beta |^2
\right.
}
\nonumber\\&&\hspace{2ex}
\left.
-\,B_{\rm out}\left(\beta ^2 +  \beta ^{\ast 2}\right)
+ C_{\rm out}^\ast\beta  + C_{\rm out}\beta^\ast\right],
\end{eqnarray}
where
\begin{eqnarray}
\label{3.7}
&&N_{\rm out} =
\frac{2}{\lambda ^2 \sqrt{1+8\sigma\cosh \!\left(2\zeta _0\right)
+ 16\sigma ^2}} \\
&& \times
\exp\!\left\{ - \displaystyle\frac{2 |\alpha _0|^2
\left[ \cosh \!\left(2\zeta _0\right) + 4\sigma \right] +
\left( \alpha _0^2 + \alpha _0^{*2} \right) \sinh \!\left(2\zeta _0\right) }
{1+ 8 \sigma \cosh \!\left(2\zeta _0\right) + 16 \sigma ^2} \right\},
\nonumber
\end{eqnarray}
\begin{equation}
\label{3.8}
A_{\rm out} = \frac{2\left[\cosh \!\left(2\zeta _0\right)
+ 4\sigma\right]}{\lambda ^2
\left[1+8\sigma\cosh\!\left( 2\zeta _0\right) + 16\sigma ^2 \right]}\,,
\end{equation}
\begin{equation}
\label{3.9}
B_{\rm out} = \frac{\sinh \!\left(2\zeta _0\right)}{\lambda ^2
\left[1+8\sigma\cosh \!\left(2\zeta _0\right) + 16\sigma ^2 \right]}\,,
\end{equation}
\begin{equation}
\label{3.10}
C_{\rm out} =
2\frac{ \alpha _0 \left[ \cosh \!\left(2\zeta _0\right) + 4\sigma \right] +
\alpha _0^* \sinh \!\left(2\zeta _0\right) }
{\lambda \left[ 1+ 8 \sigma \cosh \!\left(2\zeta _0\right)
+ 16 \sigma ^2 \right] }
\end{equation}
($\tilde{\varphi}$ $\!=$ $\!0$). Combining Eqs.~(\ref{2.18}),
(\ref{3.1}) -- (\ref{3.5}), and (\ref{3.6}) -- (\ref{3.10}),
we arrive at the following
expression for the fidelity (Appendix \ref{app1}):
\begin{eqnarray}
\label{3.11}
\lefteqn{
F \equiv F(\zeta _0,\alpha _0) =
F(\zeta _0)
\exp\! \left\{ -\frac{(1-\lambda )^2}{2} \right.
}
\nonumber \\ && \times
\left.
\left[
\frac{ \left( \alpha _0 + \alpha _0^{*} \right) ^2
e^{2\zeta _0} }{1+\lambda ^2
\left(1 + 4 e^{2\zeta _0} \sigma \right)}
- \frac{ \left( \alpha _0 - \alpha _0^{*} \right) ^2 }
{\left(1 + \lambda ^2 \right) e^{2\zeta _0}
+ 4\lambda ^2 \sigma }
\right] \right\},
\end{eqnarray}
where
\begin{eqnarray}
\label{3.12}
\lefteqn{
F(\zeta _0) = 2\left[
1 + 2 \lambda ^2 + \lambda ^4 \left(1
+ 16 \sigma ^2 \right)
\right.
}
\nonumber \\[.5ex] &&\hspace{9ex}
\left.
+\, 8 \lambda ^2
\left(1 + \lambda ^2 \right) \sigma
\cosh \!\left(2\zeta _0\right) \right]^{-\frac{1}{2}}
\end{eqnarray}
is the fidelity for teleporting the squeezed vacuum.

{F}rom Eq.~(\ref{3.11}) it is seen that the dependence on
$\alpha _0$ of $F$ vanishes for \mbox{$\lambda$ $\!=$ $\!1$}.
Thus, the fidelity of teleportation of a squeezed coherent state
can only depend on the coherent amplitude for an asymmetrical
equipment (i.e., \mbox{$|T_1|$ $\!\neq$ $\!|T_2|$}).
In this case, the fidelity exponentially decreases with
increasing coherent amplitude.
For stronger squeezing of the signal mode, the effect is more
pronounced for amplitude squeezing [first term in the
square brackets in the exponential in Eq.~(\ref{3.11})]
than for phase squeezing (second term in the square brackets).

In the case of a squeezed state, the characteristic scale
$\delta_W$ in the inequality (\ref{2.25}) is of the order
of magnitude of the small semi-axis of the squeezing ellipse,
\begin{equation}
\label{3.13}
\delta_W \sim
e^{-|\zeta _0|}.
\end{equation}
For \mbox{$|T_1|$ $\!\approx$ $\!|T_2|$ $\!=$ $\!|T|$},
from Eqs.~(\ref{2.20})
and (\ref{3.13}) it then follows that the condition
(\ref{2.25}) for high-fidelity teleportation
corresponds to
\begin{equation}
\label{3.14}
1 - |T|^2 \ll e^{-2|\zeta_0|},
\end{equation}
that is,
\begin{equation}
\label{3.15}
\frac{l}{l_{\rm A}} \ll
\ln\!\left(1 - e^{-2|\zeta_0|} \right)^{-\frac{1}{2}}.
\end{equation}
Thus, for large values of the squeezing parameter $|\zeta_0|$, the
largest teleportation distance that is possible, $l_{\rm T}$,
scales as
\begin{equation}
\label{3.16}
l_{\rm T} \sim l_{\rm A}\, e^{-2|\zeta_0|}.
\end{equation}

%%%%%%%%%%%%%%%%%%%%%%%%%%%%%%%%%%%%%%%%%%%%%%%%%%%%%%%%%%%%%%%%%%%%%%%%

\subsection{Number states}
\label{fock}

Let us now consider the case when the unknown quantum state is an
$N$-photon number state. The input Wigner function then reads
\begin{equation}
\label{wfock}
W_{\rm in}(\gamma ) = (-1)^N \frac{2}{\pi} \,
\exp\!\left( -2|\gamma |^2\right)
{\rm L}_N\!\left(4|\gamma |^2\right)
\end{equation}
[${\rm L}_N(x)$, Laguerre polynomial]. We substitute this expression
into Eq.~(\ref{2.22}) and derive the Wigner function of the
teleported state as
\begin{eqnarray}
\label{wfk}
\lefteqn{
W_{\rm out}(\beta ) = \frac{2}{\pi \lambda ^2}
\frac{(4\sigma - 1)^N}{(4\sigma + 1)^{N+1}}
}
\nonumber\\[.5ex]&&\hspace{2ex}\times\,
\exp\!\left[ -\frac{\displaystyle 2|\beta |^2}
{\displaystyle \lambda ^2 (4\sigma + 1)} \right]
\,{\rm L}_N\!\left[ -\frac{4|\beta |^2}
{\lambda ^2 (16\sigma ^2 - 1)} \right].
\end{eqnarray}
\end{multicols}
\vspace*{10mm}
%%%%%%%%%%%%%%%%%%%%%%%%%%%%%% FIGURE 1 %%%%%%%%%%%%%%%%%%%%%%%%%%%%%%%%%%%%%
\begin{figure}[htb]
\psfig{file=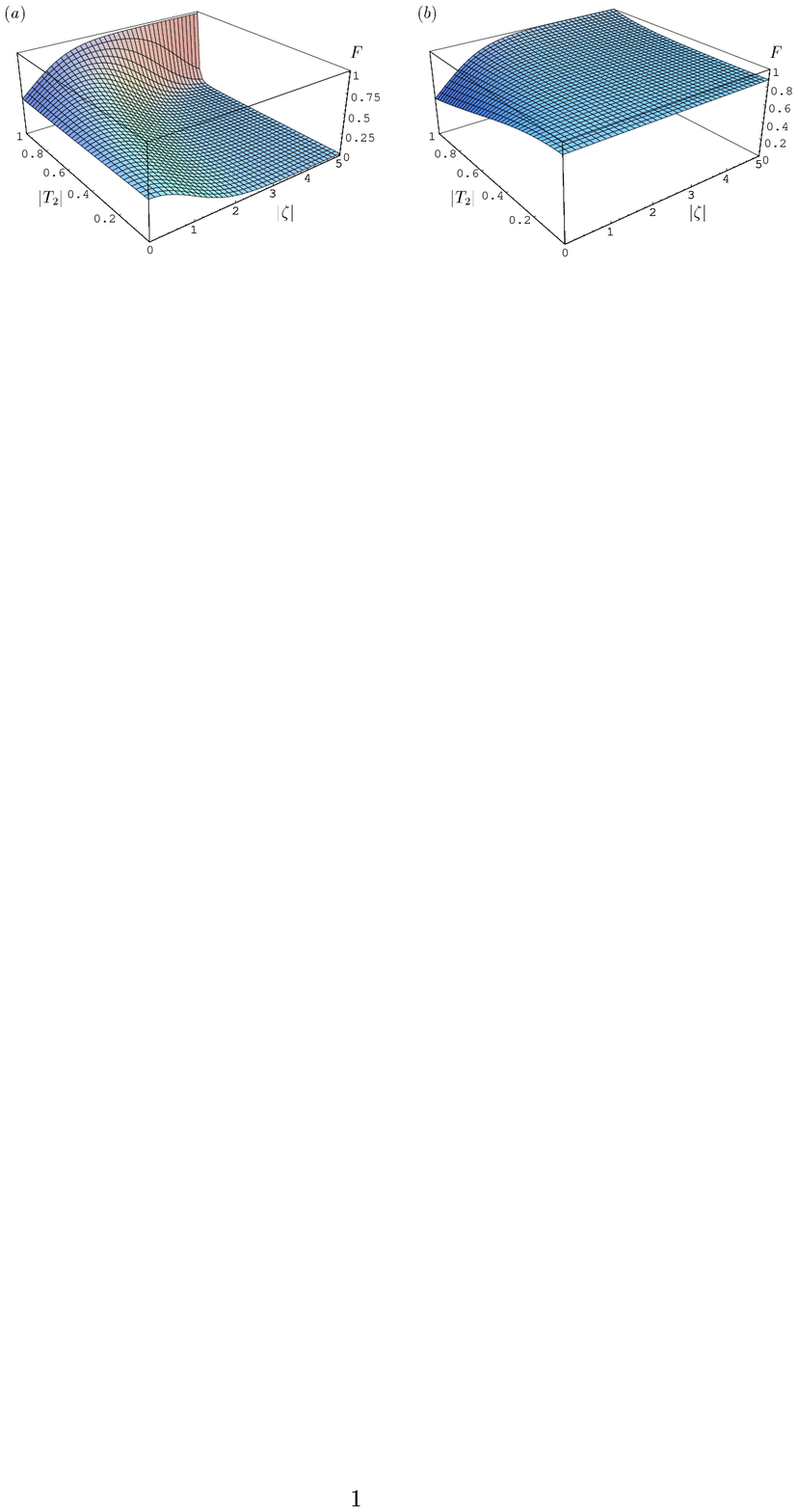,width=\textwidth} %}
\caption{\label{fig1}
The fidelity of teleportation of a squeezed vacuum state
\mbox{($\zeta_0$ $\!=$ $\!0.5$)} is shown as a function of $|\zeta|$
and $|T_2|$ (\mbox{$|T_1|$ $\!=$ $\!1$},
\mbox{$\tilde{\varphi}$ $\!=$ $\!0$}) for the displacement
$(a)$ \mbox{$\Delta(\gamma')$ $\!=$ $\!\gamma'$} and
$(b)$ \mbox{$\Delta(\gamma')$ $\!=$ $\!|T_2/T_1|\gamma'$}
[Eqs.~(\protect\ref{2.21}) and (\protect\ref{2.23})].
}
\end{figure}
%%%%%%%%%%%%%%%%%%%%%%%%%%%%%%%%%%%%%%%%%%%%%%%%%%%%%%%%%%%%%%%%%%%%%%%%%%%%%
\begin{multicols}{2}
By combining Eqs.~(\ref{2.18}), (\ref{wfock}), and
(\ref{wfk}), we then obtain the teleportation fidelity
\begin{eqnarray}
\label{fidfock}
\lefteqn{
F \equiv F_{\rm N} = 2 \frac{[\lambda ^2(4\sigma - 1)-1]^N}
{[\lambda ^2(4\sigma + 1)+1]^{N+1}}
}
\nonumber\\[.5ex]&&\hspace{2ex}\times\,
{\rm P}_N\!\left\{ 1 + \frac{8\lambda ^2}{[\lambda ^2(4\sigma +1)+1]
[\lambda ^2(4\sigma -1)-1]} \right\}
\end{eqnarray}
[${\rm P}_N(x)$, Legendre polynomial].

{F}rom inspection of Eq.~(\ref{wfock}) it is clear that
the characteristic length scale $\delta_W$ in the inequality
(\ref{2.25}) may be assumed to be of the order of magnitude
of the (difference of two neighbouring) roots of the
Laguerre polynomial ${\rm L}_N(x)$,
which for large $N$ ({$N$ $\!\gtrsim$ $\!3$}) behaves like $N^{-1}$
\cite{AbSteg}, thus
\begin{equation}
\label{restrN0}
\delta_{W} \sim \frac{1}{\sqrt{N}}\,.
\end{equation}
Assuming again \mbox{$|T_1|$ $\!\approx$ $\!|T_2|$ $\!=$ $\!|T|$},
the condition (\ref{2.25}) together with Eq.~(\ref{2.20})
and $\delta_W^2$ according to Eq.~(\ref{restrN0}) gives
\begin{equation}
\label{restrN1}
1 - |T|^2 \ll
\frac{1}{N}\,.
\end{equation}
It ensures that the oscillations of the Wigner function, which are
typically observed for a number state, are resolved. Otherwise Bob
cannot recognize the teleported state as an $N$-photon number state.
Hence, the largest teleportation distance that is possible
scales (for large $N$) as
\begin{equation}
\label{restrN2}
l_{\rm T} \sim \frac{l_{\rm A}}{N}\,.
\end{equation}

%%%%%%%%%%%%%%%%%%%%%%%%%%%%%%%%%%%%%%%%%%%%%%%%%%%%%%%%%%%%%%%%%%%%%%%%%%%%%

\subsection{Discussion}
\label{disc}

Whereas for perfect teleportation, i.e., $|T_1|$ $\!=$ $\!|T_2|$
$\!=$ $\!1$, Bob has to perform a displacement $\Delta(\gamma')$
$\!=$ $\!e^{i\varphi}\gamma'$
[Eq.~(\ref{2.21}) for \mbox{$\lambda$ $\!=$ $\!1$}],
which does not depend on the
position of the source of the TMSV, the situation drastically
changes for nonperfect teleportation. The effect is clearly seen
from a comparison of Fig.~\ref{fig1}(a) with Fig.~\ref{fig1}$(b)$.
In the two figures, the fidelity for teleporting
a squeezed vacuum state is shown as a function of the
squeezing parameter $|\zeta|$ of the TMSV and the transmission
coefficient $|T_2|$ for the case when the source of the TMSV is
in Alice's hand, i.e., $|T_1|$ $\!=$ $\!1$.
Figure \ref{fig1}$(a)$ shows the result that is obtained
for \mbox{$\Delta(\gamma')$ $\!=$ $\!\gamma'$}.
It is seen that when $|T_2|$ is not close to unity, then the
fidelity reduces, with increasing $|\zeta|$, below the classical
level (realized for \mbox{$|\zeta|$ $\!=$ $\!0$}). In contrast,
the displacement $\Delta(\gamma')$ $\!=$ $\!e^{i\tilde{\varphi}}
\lambda \gamma'$ with $\lambda$ from Eq.~(\ref{2.23}) ensures
that the fidelity exceeds the
classical level [Fig.~\ref{fig1}$(b)$].

\begin{figure}[t]
\psfig{file=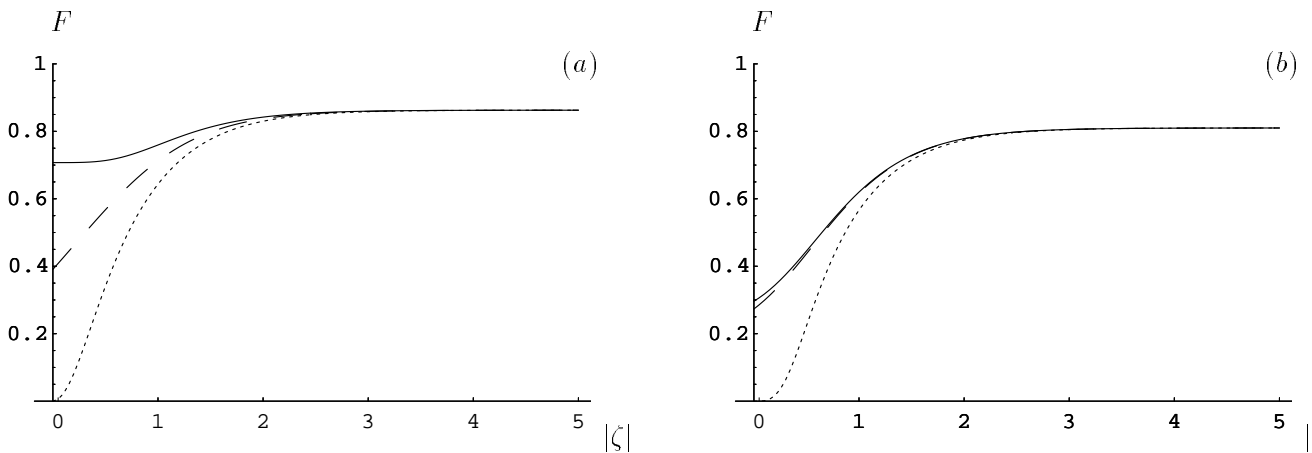,width=\textwidth}
\caption{\label{fig2}
The fidelity of teleportation of $(a)$ a squeezed vacuum state
(\mbox{$\zeta_0$ $\!=$ $\!0.88$}, i.e., \mbox{$\bar{n}$ $\!\approx$ $\!1$})
and $(b)$ a single-photon number state (\mbox{$N$ $\!=$ $\!1$})
is shown as a function of $|\zeta|$ (\mbox{$|T_1|$ $\!=$ $\!1$},
\mbox{$|T_2|$ $\!=$ $\!0.9$}, \mbox{$\tilde{\varphi}$ $\!=$ $\!0$}).
The parameter $\lambda$ in the displacement
\mbox{$\Delta(\gamma')$ $\!=$ $\!\lambda\gamma'$} is chosen such
that maximum fidelity is realized. For comparison, the fidelities
that are realized for \mbox{$\lambda$ $\!=$ $\!|T_2/T_1|$} (dashed line)
and \mbox{$\lambda$ $\!=$ $\!C_2/|S|$} (dotted line) are shown.
}
\end{figure}

At this point the question may arise of whether is the
choice of $\lambda$ according to Eq.~(\ref{2.23}) the best one
or not. For example, from inspection of Eq.~(\ref{2.19}) it could
possibly be expected that $\lambda$ $\!=$ $\!C_2/|S|$ be also a
good choice. To answer the question, we note that in
the formulas for the teleported quantum state and the fidelity,
$\lambda$ can be regarded
as being an arbitrary (positive) parameter that must not necessarily
be given by Eq.~(\ref{2.23}). Hence for chosen signal state
and given value of $|\zeta|$, the value of $\lambda$ (and thus
the value of the displacement) that maximizes the teleportation
fidelity can be determined. Examples of the fidelity (as a function
of $|\zeta|)$ that can be realized in this way are shown in
Fig.~\ref{fig2} for teleporting squeezed and number
states according to Eqs.~(\ref{3.11}) and (\ref{fidfock}),
respectively. The figure reveals that for not too small
values of $|\zeta|$, that is, in the proper teleportation regime,
the state-{\em independent} choice of $\lambda$ according to
Eq.~(\ref{2.23}) is indeed the best one.
\end{multicols}

\vspace*{10mm}
%%%%%%%%%%%%%%%%%%%%%%%%%%%%%%%% FIGURE 2 %%%%%%%%%%%%%%%%%%%%%%%%%%%%%%%%%%%
%%%%%%%%%%%%%%%%%%%%%%%%%%%%%%%%%%%%%%%%%%%%%%%%%%%%%%%%%%%%%%%%%%%%%%%%%%%%%
\begin{multicols}{2}

After preparing this manuscript we have been aware of the article
\cite{Kim01} in which it is argued that (even in the limit of infinite
squeezing of the TMSV) the average coherent-state teleportation
fidelity, which is obtained when
integrating Eq.~(\ref{3.11}) (\mbox{$\zeta_0$ $\!=$ $\!0$})
over all coherent displacements $\alpha_0$, is maximized
for \mbox{$\lambda$ $\!=$ $\!1$}. This is certainly not correct.
To see this, let us define the average fidelity more rigorously by
introducing an appropriately chosen regularizing function,
\begin{equation}
\label{cutoff}
\bar{F} =
\frac{1}{\pi\bar{n}_{\rm coh}}
\int d^2\alpha_0 \, F(\alpha_0) e^{-|\alpha_0|^2/\bar{n}_{\rm coh}}
\end{equation}
[$F(\alpha_0)$ $\!\equiv$ $F(\zeta_0$ $\!=$ $\!0,\alpha_0 )$ with
$F(\zeta_0,\alpha_0)$ from Eq.~(\ref{3.11})],
and look (for chosen $|\zeta|$ and chosen ``cutoff'' coherent-photon
number $\bar{n}_{\rm coh}$) for the value of $\lambda$ that
maximizes $\bar{F}$.
%%%%%%%%%%%%%%%%%%%%%%%%%%%%%%% FIGURE 3 %%%%%%%%%%%%%%%%%%%%%%%%%%%%%%
\begin{figure}[h]
\begin{center}
\mbox{\psfig{file=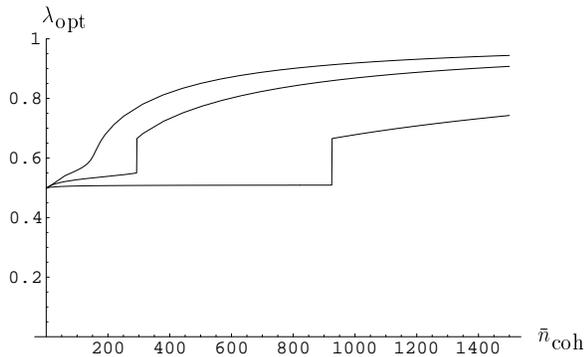,width=.45\textwidth}}
\end{center}
\caption{\label{fig3}
The optimum displacement parameter
for teleporting coherent states through
an asymmetrical equipment (\mbox{$T_1$ $\!=$ $\!1$},
\mbox{$|T_2|$ $\!=$ $\!0.5$}) is shown as a function of the
cutoff coherent-photon number $\bar{n}_{\rm coh}$. The curves
correspond to squeezing parameters
\mbox{$|\zeta|=3$} (upper curve), \mbox{$|\zeta|=3.3$} (middle curve),
and \mbox{$|\zeta|=4$} (lower curve) of the initial TMSV.}
\end{figure}
%%%%%%%%%%%%%%%%%%%%%%%%%%%%%%%%%%%%%%%%%%%%%%%%%%%%%%%%%%%%%%%%%%%%%%%
Figure~\ref{fig3} shows the optimum displacement
as a function of $\bar{n}_{\rm coh}$ for various values of
$|\zeta|$
(\mbox{$T_1$ $\!=$ $\!1$}, \mbox{$|T_2|$ $\!=$ $\!0.5$}).
One observes that for
each value of $\bar{n}_{\rm coh}$ there exists always a
(sufficiently large) value of $|\zeta|$ such that
the optimum value of $\lambda$ is exactly $|T_2/T_1|$.
Thus, when performing the limits in the
order required by quantum teleportation (first
\mbox{$|\zeta|$ $\!\to$ $\!\infty$}, then \mbox{$\bar{n}_{\rm
coh}$ $\!\to$ $\!\infty$}), the optimum displacement is always
\mbox{$\lambda_{\rm opt}$ $\!=$ $\!|T_2/T_1|$}. Note that
this is also valid when in Eq.~(\ref{cutoff})
$F(\alpha_0)$ is replaced with $F(\zeta_0,\alpha_0)$
for arbitrary $\zeta_0$. By the way, the average fidelity obtained in
\cite{Kim01} is nothing but the fidelity for teleporting the vacuum,
with not properly chosen displacement, i.e., $F(\zeta_0)$
from Eq.~(\ref{3.11}) for \mbox{$\zeta_0$ $\!=$ $\!0$}
and \mbox{$\lambda$ $\!=$ $\!1$}.

\begin{figure}[ht]
\psfig{file=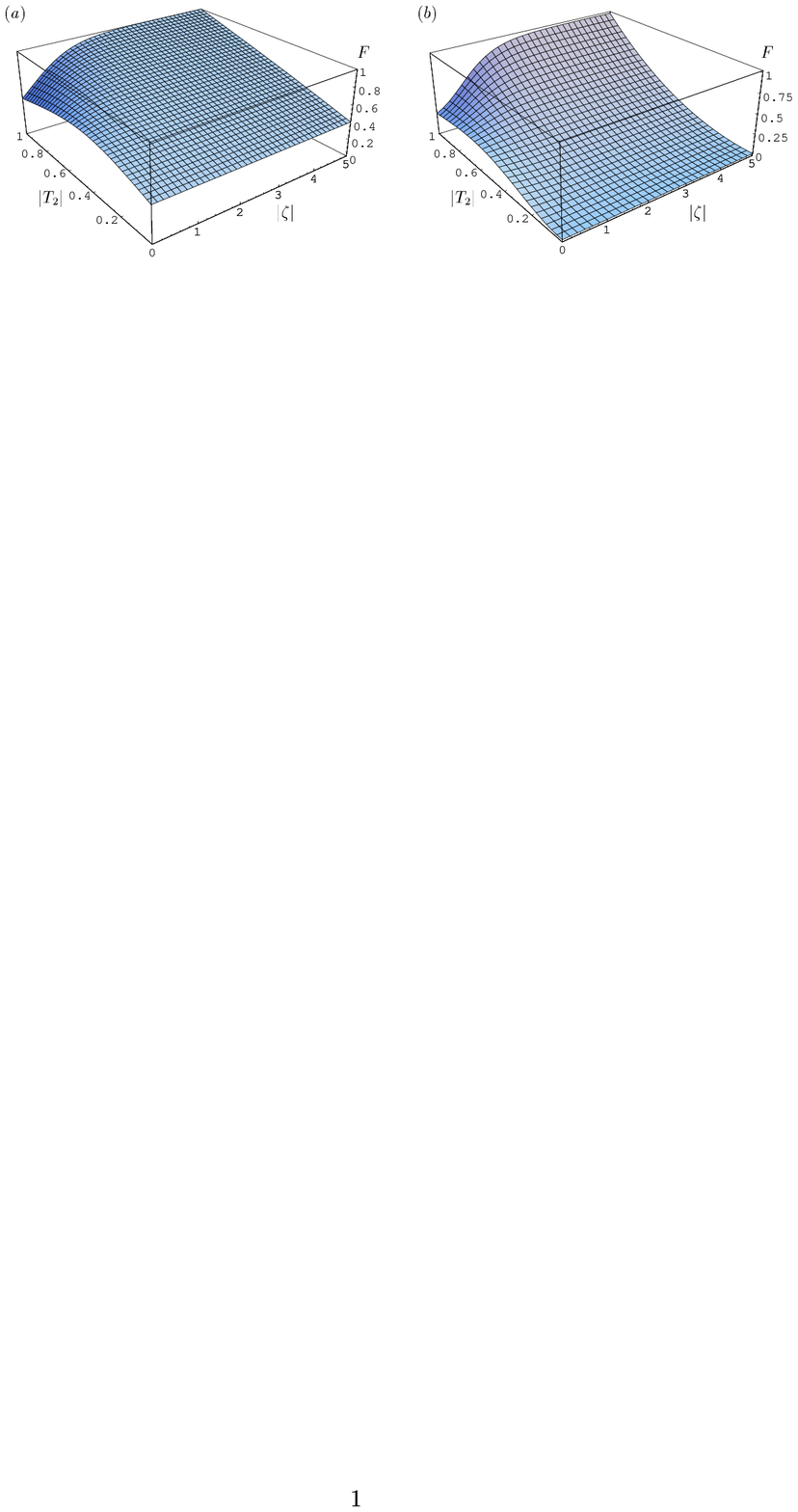,width=\textwidth}
\caption{\label{fig4}
The fidelity of teleportation of $(a)$ a squeezed coherent state
(\mbox{$\zeta_0$ $\!=$ $\!0.5$}, \mbox{$\alpha_0$ $\!\approx$ $\!0.7$},
i.e., \mbox{$\bar{n}$ $\!\approx$ $\!1$})
and $(b)$ a single-photon number state (\mbox{$N$ $\!=$ $\!1$})
is shown as a function of $|\zeta|$ and $|T_2|$
(\mbox{$|T_1|$ $\!=$ $\!1$}, \mbox{$\tilde{\varphi}$ $\!=$ $\!0$},
\mbox{$\lambda$ $\!=$ $\!|T_2/T_1|$}).
}
\end{figure}

Figure \ref{fig4} illustrates the dependence of the teleportation
fidelity on the squeezing parameter $|\zeta|$ of the TMSV and the
transmission coefficient $|T_2|$ ($|T_1|$ $\!=$ $\!1$)
for squeezed and number states.
It is seen that with increasing value of $|\zeta|$ the
fidelity is rapidly saturated below unity, because of absorption.
Even if the TMSV were infinitely entangled, the fidelity would
be noticeably smaller than unity in practice.
Only when $|T_2|$ is very close to unity, the fidelity
substantially exceeds the classical level and becomes close
to unity. (Note that the classical level is much smaller for
number states than for squeezed states.)
Hence, it seems to be principally impossible to realize quantum
teleportation over distances that are comparable with
those of classical transmission channels. The result
is not unexpected, because the scheme is based on a strongly
squeezed TMSV, which corresponds to an entangled {\em macroscopic}
(at least {\em mesoscopic}) quantum state. Clearly, such a
state decays very rapidly, so that the potencies inherent
in it cannot be used in practice.
\end{multicols}

\vspace*{10mm}
%%%%%%%%%%%%%%%%%%%%%%%%%%%%% FIGURE 4 %%%%%%%%%%%%%%%%%%%%%%%%%%%%%%%%%%%%%%
%%%%%%%%%%%%%%%%%%%%%%%%%%%%%%%%%%%%%%%%%%%%%%%%%%%%%%%%%%%%%%%%%%%%%%%%%%%%

\begin{figure}[hb]
\psfig{file=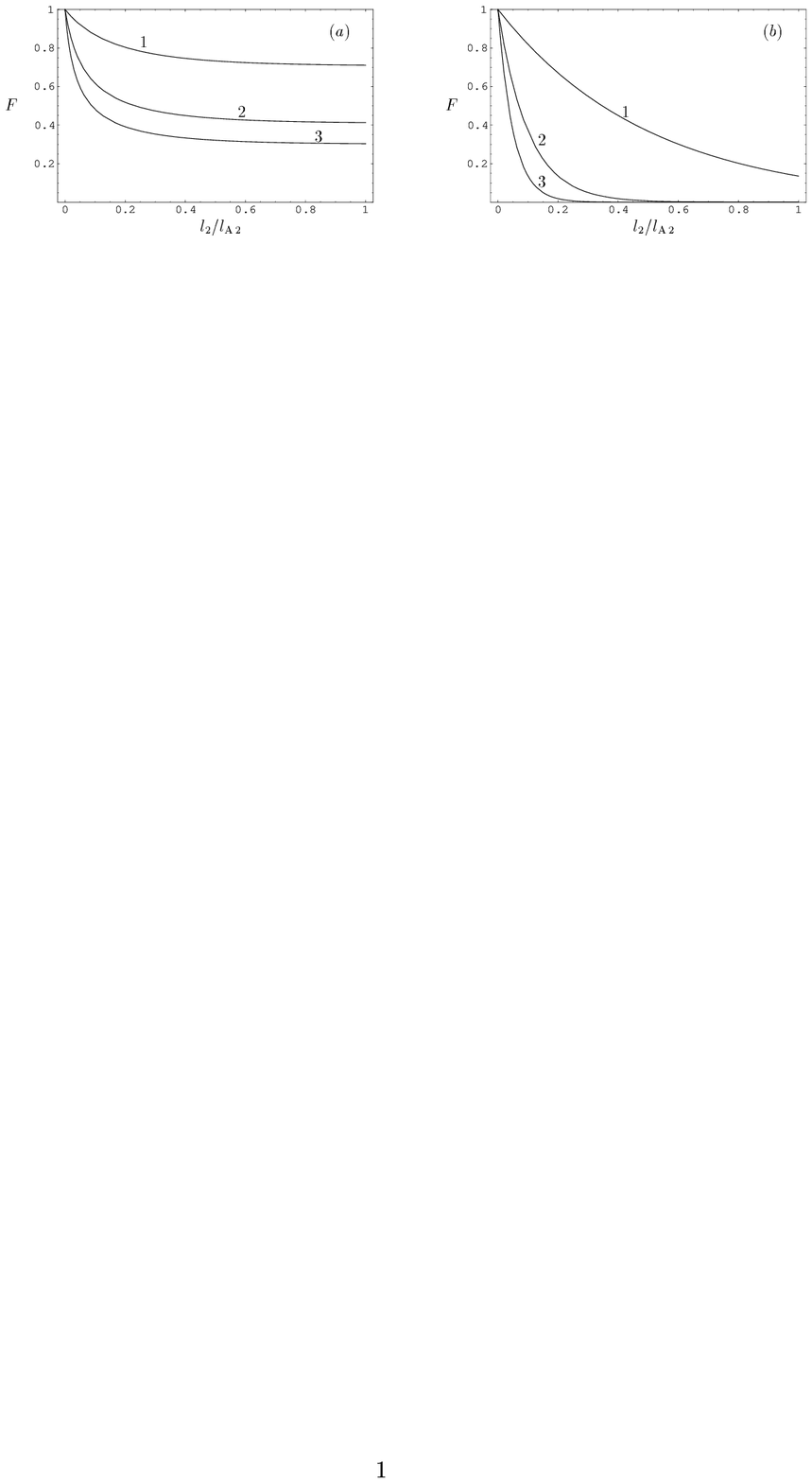,width=\textwidth}
\caption{\label{fig5}
The fidelity of teleportation of $(a)$ squeezed vacuum states
(curve $1$: \mbox{$\zeta_0$ $\!=$ $\!0.88$}, i.e.,
\mbox{$\bar{n}$ $\!\approx$ $\!1$};
curve $2$: \mbox{$\zeta_0$ $\!=$ $\!1.54$}, i.e.,
\mbox{$\bar{n}$ $\!\approx$ $\!5$};
curve $3$: \mbox{$\zeta_0$ $\!=$ $\!1.87$}, i.e.,
\mbox{$\bar{n}$ $\!\approx$ $\!10$})
and $(b)$ number states
(curve $1$: \mbox{$N$ $\!=$ $\!1$};
curve $2$: \mbox{$N$ $\!=$ $\!5$};
curve $3$: \mbox{$N$ $\!=$ $\!10$})
is shown as a function of the transmission length $l_2$
for \mbox{$|\zeta|$ $\!\to$ $\!\infty$}
[\mbox{$l_1$ $\!=$ $\!0$}, \mbox{$\tilde{\varphi}$ $\!=$ $\!0$},
\mbox{$\lambda$ $\!=$ $\!|T_2/T_1|$
$\!=$ $\!\exp (-l_2/l_{{\rm A} \, 2})$}].
}
\end{figure}

\begin{multicols}{2}
In order to illustrate the ultimate limits in more detail,
we have plotted in Fig.~\ref{fig5} the dependence of the
teleportation fidelity on the transmission length $l_2$
(\mbox{$l_1$ $\!=$ $\!0$}) for squeezed and number
states, assuming an infinitely squeezed TMSV. It is seen
that with increasing transmission length the fidelity very
rapidly decreases, and it approaches the classical level on a
length scale that is much shorter than the absorption length.
In particular, the distance over which a squeezed state can really
be teleported drastically decreases with increasing squeezing.
Exactly the same effect is observed for number states
when the number of photons increases.
In other words, for chosen distance, the amount of information
that can be transferred quantum mechanically is limited,
so that essential information about the quantum state that is
desired to be teleported may be lost.

\begin{figure}[h]
\begin{center}
\mbox{\psfig{file=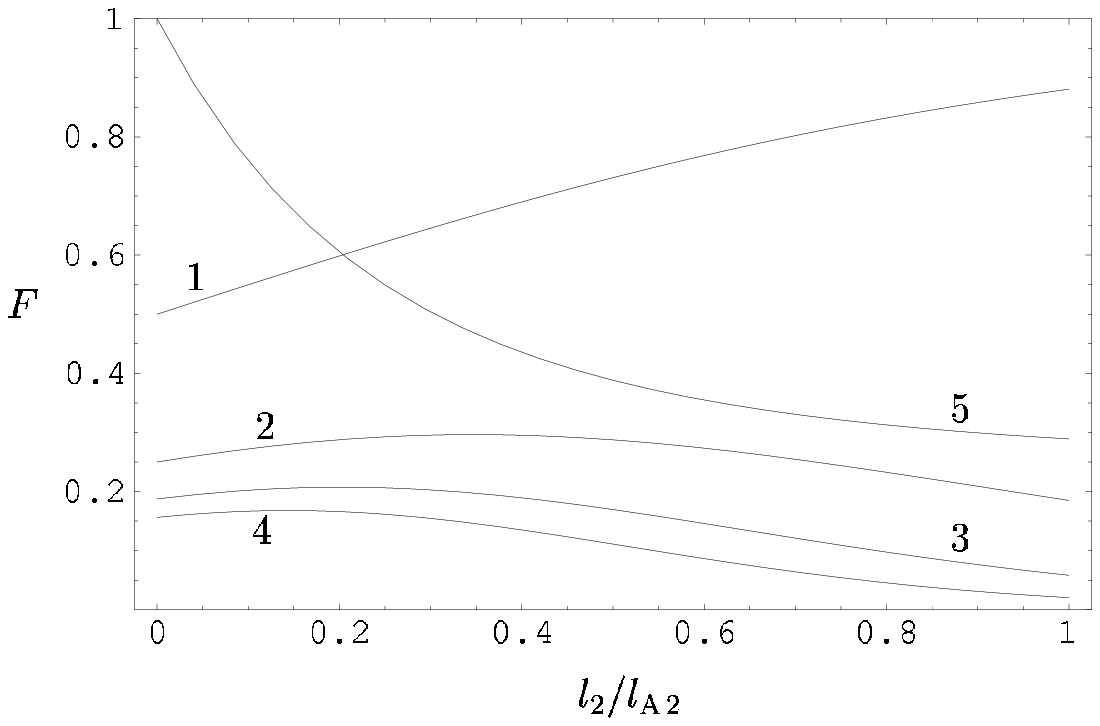,width=.45\textwidth}}
\end{center}
\caption{\label{fig6}
The classical level (\mbox{$\zeta$ $\!=$ $\!0$})
of teleporting number states (curve $1$: \mbox{$N$ $\!=$ $\!0$};
curve $2$: \mbox{$N$ $\!=$ $\!1$}; curve $3$: \mbox{$N$ $\!=$ $\!2$};
curve $4$: \mbox{$N$ $\!=$ $\!3$}) and the corresponding average
fidelity (\mbox{$|\zeta|$ $\!\to$ $\!\infty$}, curve $5$)
are shown as functions of the transmission length $l_2$
[\mbox{$l_1$ $\!=$ $\!0$}, \mbox{$\tilde{\varphi}$ $\!=$ $\!0$},
\mbox{$\lambda$ $\!=$ $\!|T_2/T_1|$
$\!=$ $\!\exp (-l_2/l_{{\rm A} \, 2})$}].
}
\end{figure}

It may be interesting to compare the (maximally realizable)
teleportation fidelity (\mbox{$|\zeta|$ $\!\to$ $\!\infty$}) with
the classical level (\mbox{$\zeta$ $\!=$ $\!0$}). Figure
\ref{fig6} illustrates the dependence of the classical level
on the transmission length $l_2$ (\mbox{$l_1$ $\!=$ $\!0$})
for the number states $|0\rangle\ldots|3\rangle$. For
comparison, the average fidelity for the set of states is shown.
With increasing transmission length the average fidelity rapidly
approaches the classical level, i.e., the average of the
classical levels of the set of states. In particular, it is seen
that the long-distance average fidelity is substantially
determined by the (classical level of the) vacuum teleportation.
Clearly, the vacuum state is the only state that (for the chosen
optimum displacement) can be teleported perfectly, without
doing nothing.

\begin{figure}[hb]
\psfig{file=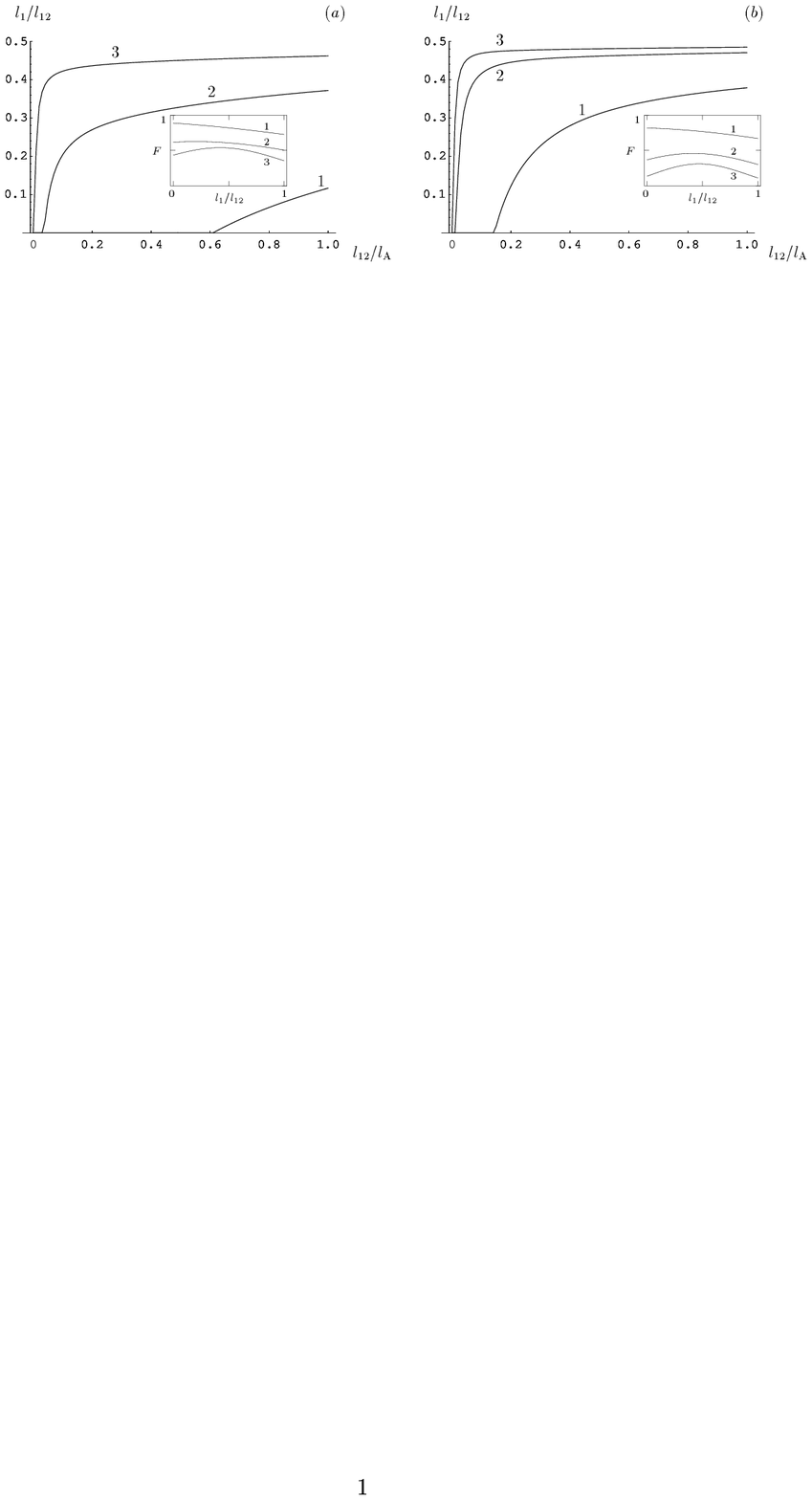,width=\textwidth}
\caption{\label{fig7}
The optimal distance $l_1$ from Alice to the position of an
infinitely squeezed TMSV, for which maximum teleportation fidelity
is realized, is shown as a function of the teleportation
distance \mbox{$l_{12}$
$\!=$ $\!l_1$ $\!+$ $\!l_2$}
(\mbox{$\tilde{\varphi}$ $\!=$ $\!0$},
\mbox{$\lambda$ $\!=$ $\!|T_2/T_1|$
$\!=$ $\!\exp [(l_1-l_2)/l_{\rm A} ]$})
for $(a)$ squeezed states
(curve $1$: \mbox{$\zeta_0$ $\!=$ $\!0.78$},
\mbox{$\alpha_0$ $\!=$ $\!0.5$}, i.e.,
\mbox{$\bar{n}$ $\!\approx$ $\!1$};
curve $2$: \mbox{$\zeta_0$ $\!=$ $\!1.44$},
\mbox{$\alpha_0$ $\!=$ $\!1$}, i.e.,
\mbox{$\bar{n}$ $\!\approx$ $\!5$};
curve $3$: \mbox{$\zeta_0$ $\!=$ $\!1.63$},
\mbox{$\alpha_0$ $\!=$ $\!2$}, i.e.,
\mbox{$\bar{n}$ $\!\approx$ $\!10$})
and $(b)$ number states
(curve $1$: \mbox{$N$ $\!=$ $\!1$};
curve $2$: \mbox{$N$ $\!=$ $\!5$};
curve $3$: \mbox{$N$ $\!=$ $\!10$}). The insets show the
dependence of the fidelity on the position of the source of
the TMSV for \mbox{$l_{12}/l_{\rm A}$ $\!=$ $\!0.1$}.
}
\end{figure}

So far we have considered the extremely asymmetrical equipment
where the source of the TMSV
is in Alice' hand (\mbox{$|T_1|$ $\!=$ $\!1$}, i.e.,
\mbox{$l_1$ $\!=$ $\!0$}).
Whereas for perfect teleportation the source of the
TMSV can be placed anywhere, in practice the teleportation fidelity
sensitively depends
\end{multicols}
%%%%%%%%%%%%%%%%%%%%%%%%%%% FIGURE 5 %%%%%%%%%%%%%%%%%%%%%%%%%%%%%%%%%%%%%%%%
%%%%%%%%%%%%%%%%%%%%%%%%%%%%%%%%%%%%%%%%%%%%%%%%%%%%%%%%%%%%%%%%%%%%%%%%%%%%
\begin{multicols}{2}
%%%%%%%%%%%%%%%%%%%%%%%%%%%%%%%%%% FIGURE 6 %%%%%%%%%%%%%%%%%%%%%%%%%%%%%%%%
%%%%%%%%%%%%%%%%%%%%%%%%%%%%%%%%%%%%%%%%%%%%%%%%%%%%%%%%%%%%%%%%%%%%%%%%%%%%
\noindent
on the position of the source of the TMSV.
In Fig.~\ref{fig7},
examples of the optimal distance
$l_1$ from Alice to the source of an infinitely squeezed TMSV
(i.e., the distance for which maximum fidelity is realized) is shown,
again for squeezed and number states, as a function of the
distance \mbox{$l_{12}$ $\!=$ $\!l_1$ $\!+$ $\!l_2$} between Alice
and Bob. It is seen that the optimal
position of the source of the TMSV is state-dependent, and it is
always nearer to Alice than to Bob (\mbox{$0$ $\!\le$ $\!l_1$
$\!<$ $\!0.5\,l_{12}$}). With increasing value of $l_{12}$
the value of $l_1$ approaches $0.5\,l_{12}$, and one could thus
think that a symmetrical equipment would be the best one.
Unfortunately, this is not the case, because the transmission
lengths are essentially too large for true quantum teleportation.
What were (optimally) observed would be the classical level at best.

%%%%%%%%%%%%%%%%%%%%%%%%%%%%%%%%%%%%%%%%%%%%%%%%%%%%%%%%%%%%%%%%%%%%%
%%%%%%%%%%%%%%%%%%%%%%%%%%%%%%%%%%%%%%%%%%%%%%%%%%%%%%%%%%%%%%%%%%%%%

\section{Summary and conclusions}
\label{sec4}

In continuous-variable single-mode quantum teleportation it
is commonly assumed that Alice and Bob share a strongly
squeezed TMSV. We have analysed this scheme, with special emphasis
on the absorption losses that are unavoidably associated with
the transmission of the two modes over finite distances,
e.g., by means of fibers. In particular, we have applied
the general formulas derived to the problem of teleporting
squeezed states and number states, which are typical
examples of nonclassical states.

The results show that the TMSV state as an effectively
macroscopic (at least mesoscopic) entangled quantum state
rapidly decays, and thus proper quantum teleportation is
only possible over distances that are much shorter than the
(classical) absorption length. Rapid decay of the TMSV state
means that there is a strong entanglement degradation which
dramatically limits the amount of information that can
be transferred quantum mechanically over longer
distances. Because of this limitation, quantum teleportation
becomes state-dependent, that is, without additional knowledge
of the state that is desired to be teleported over some
finite distance it is principally impossible to decide whether
the teleported state is sufficiently close to the original state.
Clearly, this contradicts the basic idea of quantum teleportation.
It is worth noting that both the coherent displacement that
must be performed by Bob and the position of the source of
the TMSV should not be chosen independently of the fiber lengths.
In particular, an asymmetrical equipment, where the source of
the TMSV is placed nearer to Alice than to Bob, is  suited
for realizing the largest possible teleportation fidelity
and not a symmetrical one.
\end{multicols}
\vspace*{10mm}
%%%%%%%%%%%%%%%%%%%%%%%%% FIGURE 7 %%%%%%%%%%%%%%%%%%%%%%%%%%%%%%%%%%%%%%%%%
%%%%%%%%%%%%%%%%%%%%%%%%%%%%%%%%%%%%%%%%%%%%%%%%%%%%%%%%%%%%%%%%%%%%%%%%%%%%%
\begin{multicols}{2}

To overcome the problem of fast entanglement degradation
of the TMSV, one could think about application of appropriate
purification of Gaussian continuous variable quantum states.
Unfortunately a practically realizable purification scheme
that compensates for the entanglement degradation of the TMSV
has not been known so far.
The purification protocol proposed in \cite{Duan00} enables
one to distill maximally entangled states from a mixed
two-mode entangled Gaussian state, but these states would be
far from a TMSV needed for the teleportation scheme under
consideration. In fact, the distilled states live
in finite-dimensional Hilbert spaces. Even if they could
be used in some modified teleportation scheme, the dimension
of the Hilbert space should be sufficiently high. However, since
the distillation probability exponentially decreases with
the Hilbert space dimension, one would effectively be left
with the starting problem.

Throughout this paper we have restricted our attention to
(quasi-)monochromatic fields. Using wave packets, the
ultimate limits of quantum teleportation are not only determined
by absorption but also  by dispersion. Due to dispersion, the
two wave packets unavoidably change their forms during propagation
over longer distances, and the problem of mode mismatching
in Alice's homodyne measurement and Bob's coherent displacement
appears. The corresponding quantum efficiencies
diminish, and hence the width of the Gaussian with which
the Wigner function of the original quantum state is convolved
is effectively increased. As a result, the teleportation
fidelity is reduced. It can be expected that the effect
sensitively depends on the position of the source of the TMSV.
In order to understand the details, a separate analysis is
required, which will be given elsewhere.

%%%%%%%%%%%%%%%%%%%%%%%%%%%%%%%%%%%%%%%%%%%%%%%%%%%%%%%%%%%%%%%%%%%%%
%%%%%%%%%%%%%%%%%%%%%%%%%%%%%%%%%%%%%%%%%%%%%%%%%%%%%%%%%%%%%%%%%%%%%%

\acknowledgements
This work was supported by the Deutsche Forschungsgemeinschaft,
the RFBR-BRFBR Grant No.~00-02-81023~Bel2000\_a, and the Heisenberg--Landau
Program. One of us (D.G.W.) would like to acknowledge discussions with
M. S. Kim.

%%%%%%%%%%%%%%%%%%%%%%%%%%%%%%%%%%%%%%%%%%%%%%%%%%%%%%%%%%%%%%%%%%%%%%%
%%%%%%%%%%%%%%%%%%%%%%%%%%%%%%%%%%%%%%%%%%%%%%%%%%%%%%%%%%%%%%%%%%%%%%%

\appendix
\section{Gaussian states}
\label{app1}

Let us consider the Gaussian Wigner function
\begin{eqnarray}
\label{wgin}
\lefteqn{
W_{\rm in}(\gamma ) =\frac{N_{\rm in}}{\pi }
\,\exp\bigl( - A_{\rm in}|\gamma |^2
-B_{\rm in}\gamma ^{*2}-B_{\rm in}^*\gamma ^2
}
\nonumber\\[.5ex]&&\hspace{10ex}
+\, C_{\rm in}\gamma ^* + C_{\rm in}^*\gamma
- D_{\rm in} \bigr),
\qquad
\end{eqnarray}
where
\begin{equation}
\label{wgin1}
N_{\rm in} = \sqrt{A_{\rm in}^2 -4|B_{\rm in}|^2}\,,
\end{equation}
\begin{equation}
\label{wgin2}
D_{\rm in}=\frac{1}{N_{\rm in}^2}
\left[ A_{\rm in}|C_{\rm in}|^2
- \left( B_{\rm in}^*C_{\rm in}^2+B_{\rm in}C_{\rm in}^{*2}\right)\right].
\end{equation}
Here, $C_{\rm in}$ can be chosen freely, and $A_{\rm in}$ and
$B_{\rm in}$ must be chosen such that
the condition \mbox{$A_{\rm in}$ $\! >$ $\!2 |B_{\rm in}|$} is satisfied.
Substituting Eq.~(\ref{wgin}) into Eq.~(\ref{2.22}) and performing
the integration, we again obtain a Gaussian:
\begin{eqnarray}
\label{wgout}
\lefteqn{
W_{\rm out}(\beta )
= \frac{N_{\rm out}}{\pi } \,\exp\bigl(- A_{\rm out}|\beta |^2
-B_{\rm out}\beta ^{*2}-B_{\rm out}^*\beta ^2
}
\nonumber\\[.5ex]&&\hspace{10ex}
+\, C_{\rm out}\beta ^* + C_{\rm out}^*\beta
- D_{\rm out} \bigr)
\qquad\qquad
\end{eqnarray}
($\tilde{\varphi}$ $\!=$ $\!0$), where
\begin{eqnarray}
\label{aout}
&& A_{\rm out}
=\frac{1}{4(\lambda \sigma )^2}\left[ 2\sigma -
\frac{A_{\rm in}+ 1/(2\sigma)}{\left[ A_{\rm in}
+ (1/2\sigma)\right] ^2 -4|B_{\rm in}|^2} \right],
\\[.5ex]
\label{bout}
&& B_{\rm out}
=\frac{1}{4(\lambda \sigma )^2}\frac{B_{\rm in}}{\left[ A_{\rm in}
+ 1/(2\sigma) \right] ^2 -4|B_{\rm in}|^2}\,,
\\[.5ex]
\label{cout}
&& C_{\rm out}=\frac{1}{2\lambda \sigma }\frac{\left[ A_{\rm in}
+ 1/(2\sigma) \right] C_{\rm in}-2B_{\rm in}C_{\rm in}^*}
{\left[ A_{\rm in}+1/(2\sigma) \right] ^2 -4|B_{\rm in}|^2}\,,
\end{eqnarray}
and $N_{\rm out}$ and $D_{\rm out}$ are given according to Eqs.~(\ref{wgin1})
and (\ref{wgin2}), respectively, with the out-quantities in place of
the in-quantities. The result is
\begin{eqnarray}
\label{nout}
\lefteqn{
N_{\rm out}
=\frac{1}{2\lambda ^2 \sigma }\frac{N_{\rm in}}
{\sqrt{\left[ A_{\rm in}
+ 1/(2\sigma)\right] ^2 -4|B_{\rm in}|^2}}\,,
}
\\[.5ex]&&
\label{dout}
D_{\rm out} = D_{\rm in}
\nonumber\\[.5ex]&&\hspace{2ex}
-\frac{\left[ A_{in}+ 1/(2\sigma)\right] |C_{\rm in}|^2
-\left( B_{\rm in}^* C_{\rm in}^2 + B_{\rm in}C_{\rm in}^{*2}\right)}
{\left[ A_{\rm in}
+ 1/(2\sigma) \right] ^2 -4|B_{\rm in}|^2}\,.
\end{eqnarray}
Specifying $A_{\rm in}$, $B_{\rm in}$,
and $C_{\rm in}$ according to Eqs.~(\ref{3.3}) -- (\ref{3.5}),
Eq.~(\ref{wgout}) [together with Eqs.(\ref{aout}) -- (\ref{dout})],
we arrive at Eq.~(\ref{3.6}) [together with Eqs.(\ref{3.7})
-- (\ref{3.10})].

Using Eqs.~(\ref{wgin}) and (\ref{wgout}),
it is not difficult to prove that
\begin{eqnarray}
\label{feg}
\lefteqn{
\pi \int d^2\beta\, W_{\rm in}(\beta) W_{\rm out}(\beta)
= \frac{2N_{\rm out}}{\sqrt{A^2-4|B|^2}}
}
\nonumber\\[.5ex]&&\hspace{2ex}\times\,
\exp\left[ \frac{A |C|^2 -
\left( B^* C^2 + B C^{*2}\right) }{A^2 - 4|B|^2} - D \right],
\end{eqnarray}
where
\begin{eqnarray}
A &=& A_{\rm in} + A_{\rm out}\,,
\\
B &=& B_{\rm in} + B_{\rm out}\,,
\\
C &=& C_{\rm in} + C_{\rm out}\,,
\\
D &=& D_{\rm in} + D_{\rm out}\,.
\end{eqnarray}
The above mentioned specification of $A_{\rm in}$, $B_{\rm in}$,
and $C_{\rm in}$ then leads to Eq.~(\ref{3.11}) [together
with Eq.~(\ref{3.12})].

%%%%%%%%%%%%%%%%%%%%%%%%%%%%%%%%%%%%%%%%%%%%%%%%%%%%%%%%%%%%%%%%%%%%%%%%%%%%%%%%

\end{multicols}

%%%%%%%%%%%%%%%%%%%%%%%%%%%%%%%%%%%%%%%%%%%%%%%%%%%%%%%%%%%%%%%%

\end{document}